\documentclass[a4paper,11pt]{article}
\usepackage{pos}

\usepackage{physics}

\usepackage{wrapfig}

\usepackage[separate-uncertainty = true]{siunitx}
\DeclareSIUnit{\depth}{\gram\per\square\centi\meter}

\newcommand{\xmax}{X_{\max}}

\newcommand{\nmu}{N_\mu}

\title{Innovative Approaches to Unravel the Shower Components' Energy Spectrum with a Single Hybrid Station}
\ShortTitle{Shower Energy Spectra with a Single Hybrid Station}

\author[a,b]{P. Assis}
\author*[a,b]{R. Concei\c{c}\~ao}
\author[a,b]{P. J. Costa}
\author[a,b]{M. Freitas}
\author[a,b]{B. S. Gonz\'alez}
\author[a,b]{B. Tom\'e}

\affiliation[a]{Departamento de F\'isica, Instituto Superior T\'{e}cnico (IST),\\ Universidade de Lisboa, Av. Rovisco Pais 1, 1049-001 Lisbon, Portugal}
\affiliation[b]{Laboratório de Instrumentação e Física Experimental de Partículas (LIP),\\ Av. Prof. Gama Pinto, 2, 1649-003 Lisbon, Portugal}

\emailAdd{ruben@lip.pt}

\abstract{Accurately measuring the energy of shower particles reaching the ground remains a challenge due to the inherent limitations of typical cosmic ray experiments. In this work, we present two experimental strategies to determine the energy spectra of the electromagnetic and muonic components of extensive air showers, leveraging a single hybrid detector station within a regular cosmic ray array. This station consists of a scintillator surface detector (SSD), a water Cherenkov detector (WCD), and Resistive Plate Chambers (RPCs), with a prototype currently being tested at the Pierre Auger Observatory.

The first approach exploits the different responses of each detector to the same particles traversing them, allowing, for the first time, the extraction of the high-energy tail of the electromagnetic spectrum and the low-energy tail of the muonic spectrum. The second strategy utilizes machine learning tools to reconstruct the direction of muons using the WCD+RPC system. By correlating this information with the reconstructed muon production depth, the muon kinematical delay can be analyzed, providing access to its energy spectrum.}

\ConferenceLogo{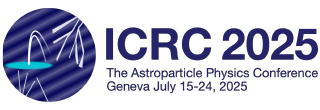}

\FullConference{39th International Cosmic Ray Conference (ICRC2025)\\
 15–24 July 2025\\
Geneva, Switzerland\\}

\begin{document}
\maketitle

\section{Introduction}
\label{sec:intro}

Ultra-high-energy cosmic rays (UHECRs) are particles with energies beyond those achievable by current accelerators. Their interactions in Earth’s atmosphere produce extensive air showers (EAS), which provide indirect insights into the properties of the primary particle and offer a unique probe of hadronic interactions at extreme energies.

Key EAS observables — such as the number of ground-level muons ($\nmu$) and the depth of shower maximum ($\xmax$) — depend on hadronic interaction models, which are uncertain at ultra-high energies due to necessary extrapolations beyond accelerator data. Recent findings from the Pierre Auger Observatory reveal a consistent discrepancy between simulations and observations, particularly a deficit in predicted muon numbers, known as the EAS Muon Puzzle~\cite{2021_Auger_muonfluctuations}.

To address these issues, efforts include both experimental upgrades like AugerPrime and new accelerator measurements (e.g., proton–oxygen collisions at the LHC), which aim to better constrain interaction models.

The concept of shower universality — the idea that shower features are determined primarily by energy and stage of development, rather than mass composition or interaction details — has been observed for both electromagnetic and muonic components. Recent studies show that hadronic models can be distinguished via the muon energy spectrum~\cite{2023_Cazon_MuonUniversality}, though the electromagnetic energy spectrum remains largely unmeasured due to design constraints of large-area cosmic ray detectors.

\begin{figure}[ht!]
    \centering
    \includegraphics[width=0.5\linewidth]{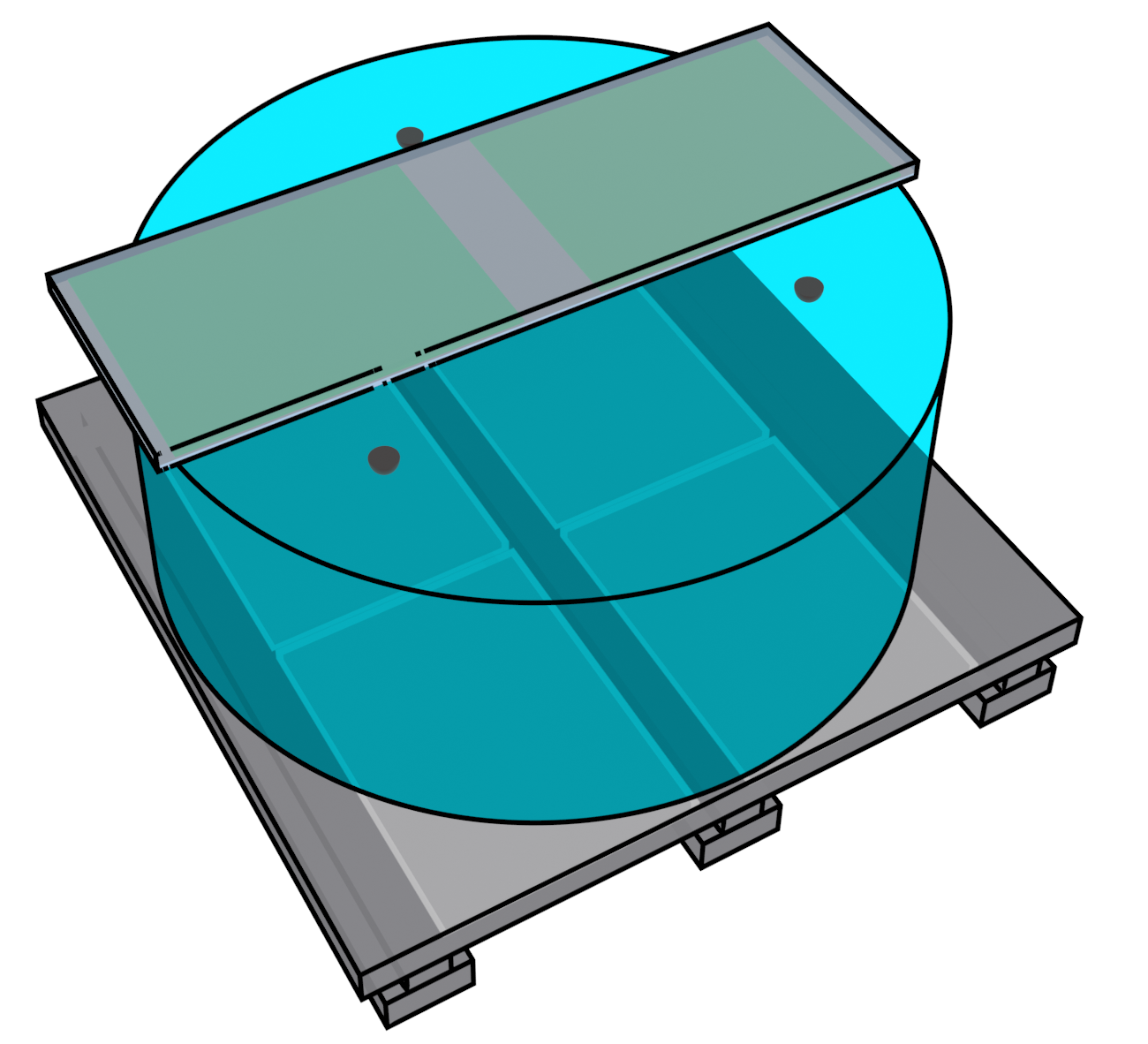}
    \caption{A schematic representation of the detector station used in this work.}
    \label{fig:scheme}
\end{figure}

In this work, we explore two approaches that use a single multi-hybrid detector configuration to probe the energy spectrum of shower secondary particles. In Sec.~\ref{sec:MARTASSD}, we combine the signals recorded by the three detectors illustrated in Fig.~\ref{fig:scheme} to demonstrate their sensitivity to the slope of the high-energy tail of the electromagnetic component of the shower at ground level. Then, in Sec.~\ref{sec:MPD}, we focus on only two of the detectors—the Water Cherenkov Detector and the Resistive Plate Chambers—to reconstruct the direction of detected muons. This directional information is used in conjunction with the Muon Production Depth (MPD) algorithm to extract the so-called kinematical delay term, which is sensitive to the energy spectrum of shower muons.

\section{Energy spectrum from a multi-hybrid station}
\label{sec:MARTASSD}

This paper presents a strategy to infer the energy spectrum of shower secondary particles by combining signals from a hybrid detector station composed of three complementary detectors. A surface scintillation detector (SSD), relatively insensitive to particle energy, is placed above a water-Cherenkov detector (WCD), which is highly energy-sensitive. To enhance calibration robustness, a third detector— resistive plate chambers (RPCs)—are positioned below the WCD, providing an independent measurement. This configuration allows particles to traverse all three detectors and is currently being tested at the Pierre Auger Observatory using the AugerPrime scintillator~\cite{2019_Castellina_AugerPrime} and the MARTA prototype station~\cite{MARTA}.

\begin{figure}[ht!]
\centering        
\includegraphics[width=0.6\linewidth]{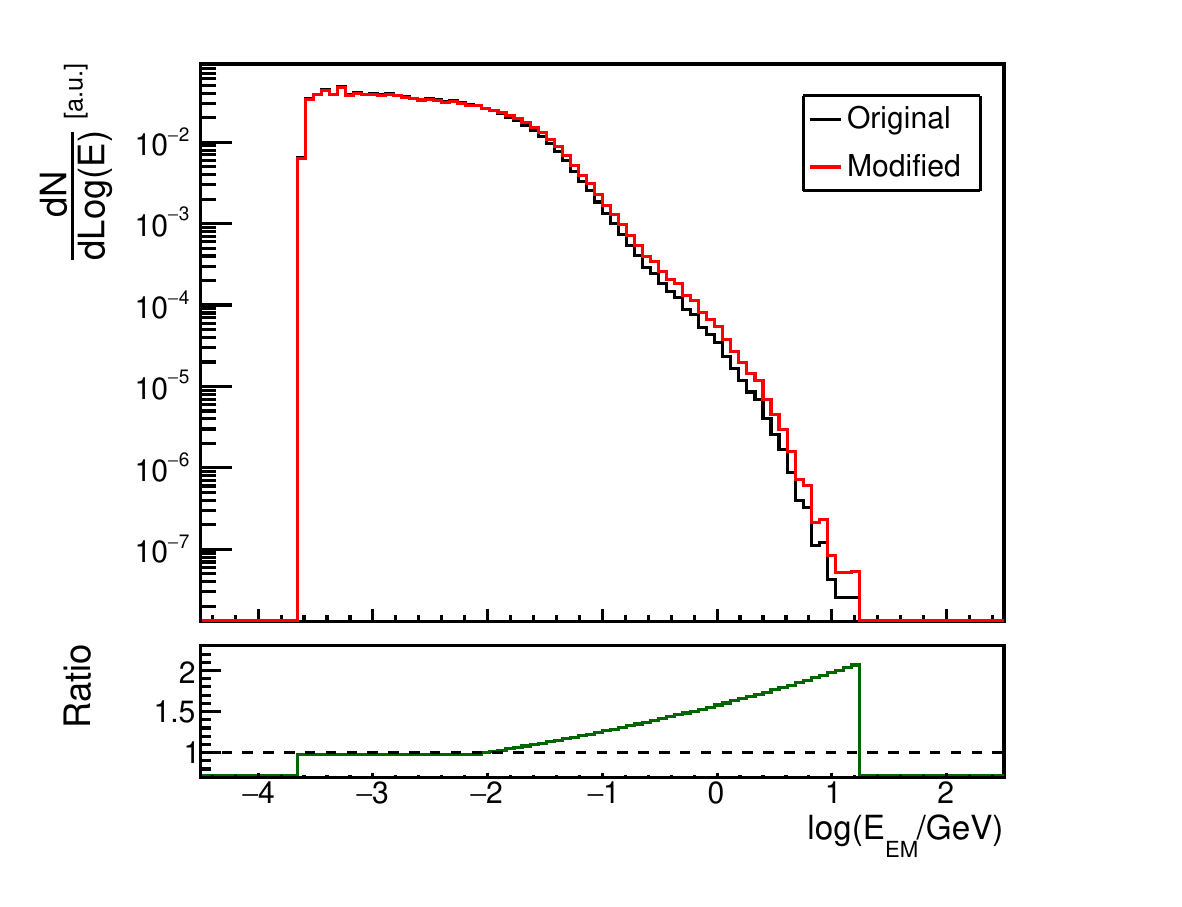}
    \caption{Example of modified energy distributions of shower secondary particles for the electromagnetic component.}
    \label{fig:spectrum}
\end{figure}

The analysis is performed at the station level rather than the shower event level to fully exploit the capabilities of the multi-hybrid detector station. This approach enables a separate evaluation of the electromagnetic and muonic components of the shower by selecting stations based on their distance from the shower core. Stations near the core are mainly influenced by the electromagnetic component, while those farther away are dominated by muons. For the next case study, we select a station located near the shower core, at a distance of \( r = 320\,\mathrm{m} \), and investigate its sensitivity to the high-energy tail of the electromagnetic shower secondaries—namely photons, electrons, and positrons—at ground level, as shown in Fig.~\ref{fig:spectrum}.

In this work, extensive air showers were simulated using \textsc{CORSIKA}~\cite{CORSIKA}, with detector responses modeled via a standalone program based on the \textsc{Geant4} toolkit~\cite{2023_Agostinelli_geant4}. A reference configuration—proton-induced showers with fixed energy \( E = 10^{17.5} \,\mathrm{eV} \) and zenith angle \( \theta = 30^\circ \)—was used, along with additional datasets to test the method's robustness in reconstructing key shower parameters. Hadronic interactions were simulated using \textsc{FLUKA}~\cite{2014_Bohlen_fluka} and \textsc{EPOS-LHC}~\cite{2015_Pierog_eposlhc}, and a thinning level of \( \varepsilon = 10^{-6} \) was applied. Each shower core was injected at a fixed position relative to a specific detector station to serve as a control setup for the analysis.

A simple threshold-based station-level trigger is implemented, in which a station is considered active if all PMTs individually register a signal of at least \( 1.75\,\mathrm{VEM} \)~\cite{2015_Auger_PAODescription}.

\begin{figure}[ht!]
    \centering
    \includegraphics[width=0.49\linewidth]{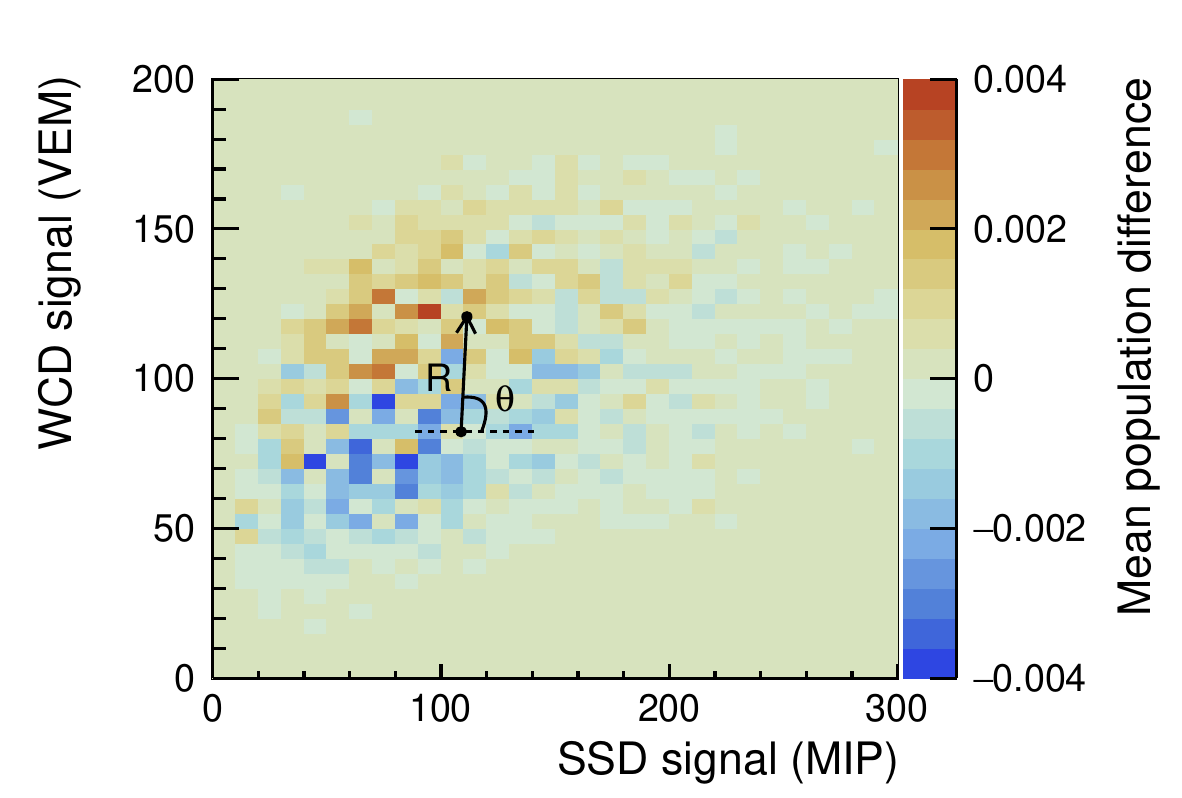}
    \includegraphics[width=0.49\linewidth]{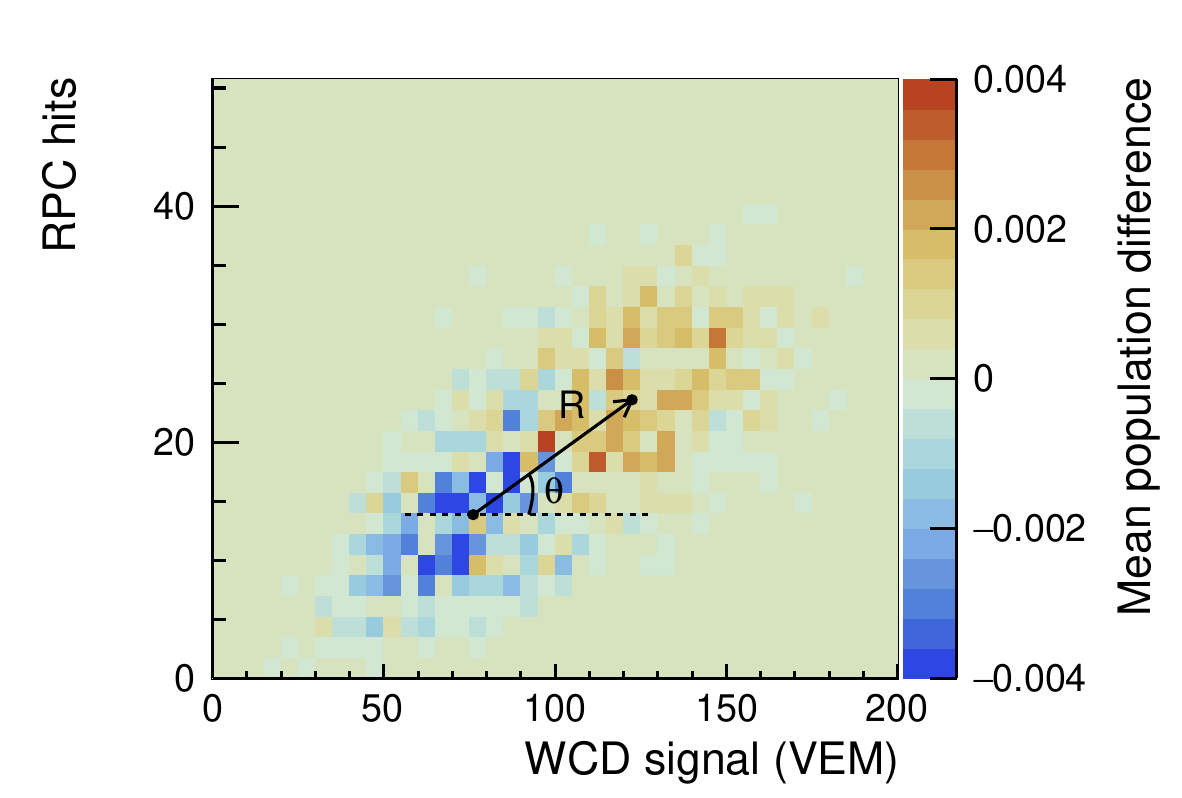}
    \caption{Modification of the high-energy tail of the electromagnetic: Difference in the SSD-WCD (left) and WCD-RPC (right) calibration plot between the modified and original distributions with a scheme illustrating the geometric representation of R and $\theta$.}
    \label{fig:diff_hists}
\end{figure}

The energy spectra of the electromagnetic and muonic components at ground level—obtained from CORSIKA simulations—are modified while preserving their overall functional shapes. This is done under the assumption that discrepancies in the spectra stem from limitations in hadronic interaction models rather than from the shower’s intrinsic physics. The method involves adjusting the slopes of the low- and high-energy tails without changing the total particle count, allowing the study of how these spectral variations affect detector response.

The modifications are applied using a smooth, energy-dependent weighting function over a selected energy range, ensuring continuity between modified and unmodified regions. After applying the weights, the distributions are renormalized to conserve particle multiplicity (see Fig.~\ref{fig:spectrum}). Particle energies are then resampled from the modified spectra, ensuring that the ensemble reflects the new distribution and allowing its impact on detector signals to be analyzed.

To highlight how modifications to the energy spectrum influence detector signal correlations, the analysis focuses on the difference between normalized signal-response histograms before and after the spectral modification, as illustrated in Fig.~\ref{fig:diff_hists}. These relative difference plots reveal qualitative shifts in the detector response distributions.

The quantification of these changes is done using two estimators, \( R \) and \( \theta \). Both are computed from the barycentres of the positive and negative regions in the relative difference plots. The estimator \( R \) corresponds to the arithmetic distance between these barycentres, while \( \theta \) denotes the angle, with respect to the \( x \)-axis, of the vector connecting them, where the vector originates at the negative barycentre. Geometric interpretations of \( R \) and \( \theta \) are provided in Fig.~\ref{fig:diff_hists} (left) and (right) for the SSD-WCD and WCD-RPC comparisons, respectively.

\begin{figure}[ht!]
    \centering
    \includegraphics[width=0.49\linewidth]{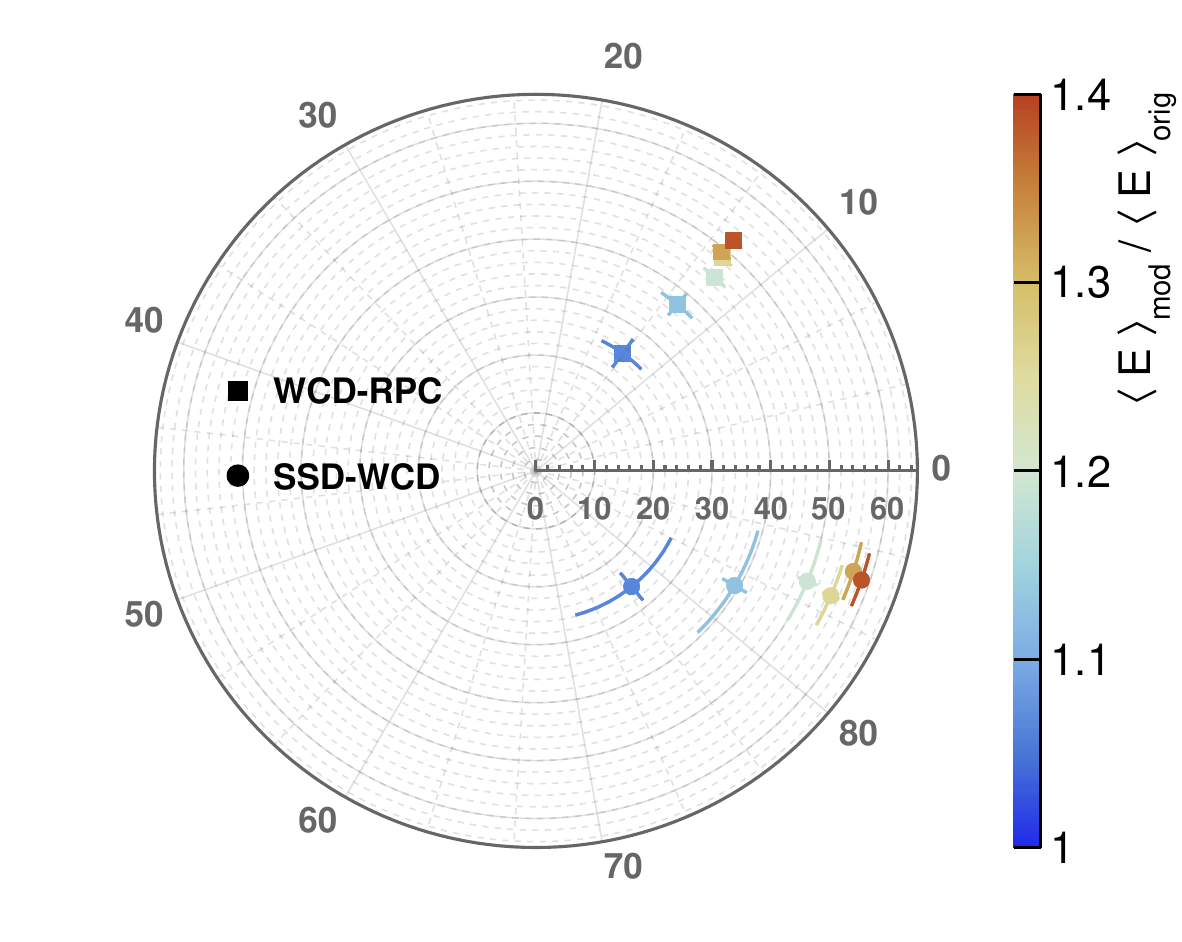}
    \includegraphics[width=0.49\linewidth]{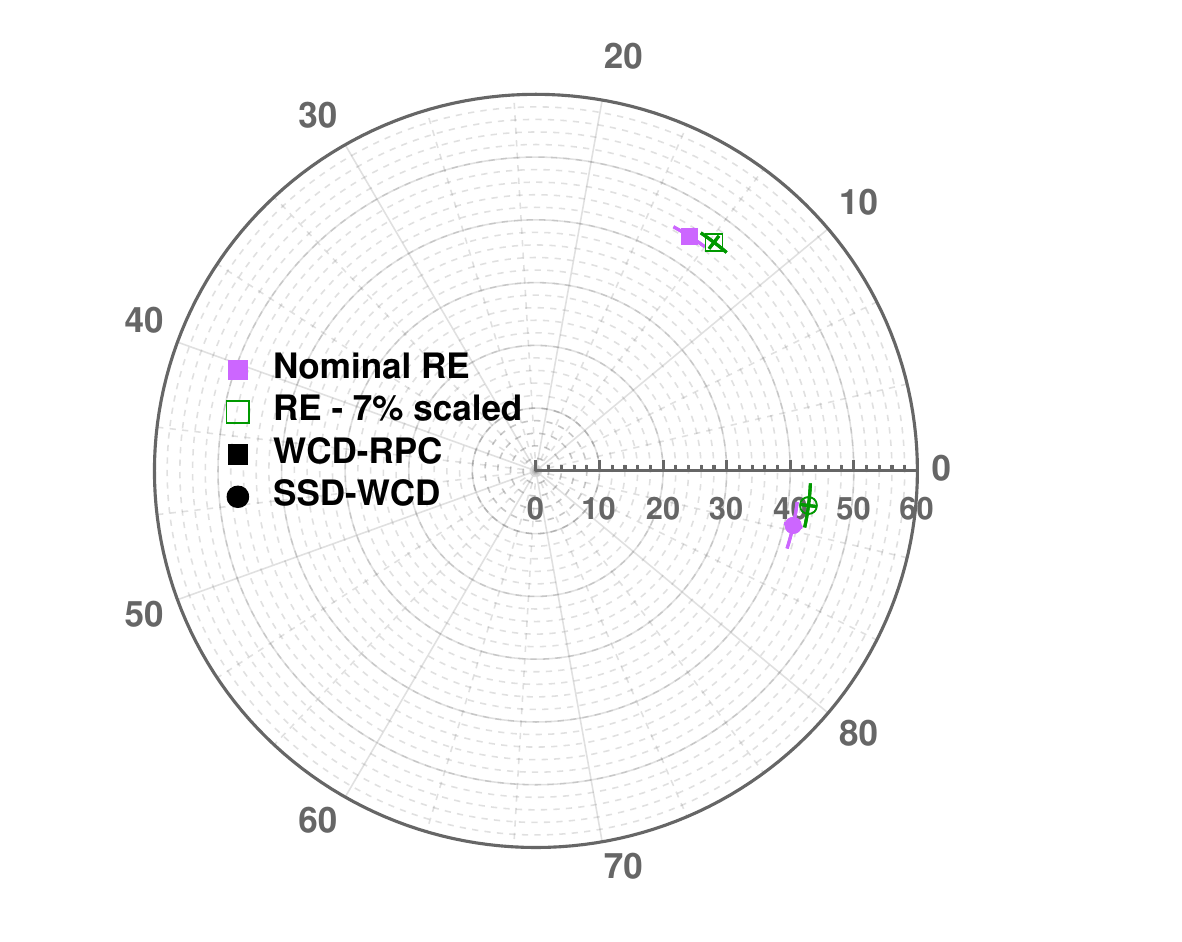}
    \caption{(left) Summary of the evolution of the  WCD-SSD and  RPC-WCD observables with the modification of the high-energy electromagnetic energy spectrum tail. (right) Summary of the dependence in changes in the reflectivity of the WCD Tyvek (see caption and text for details).}
    \label{fig:graph_polar}
\end{figure}

The results of the exercise involving variations in the high-energy tail of the electromagnetic energy spectrum are shown in Fig.~\ref{fig:graph_polar} (left). A clear dependence of the observable \( R \) on the applied spectral modification is observed. Although \( R \) and \( \theta \) do not directly measure the particle energy spectrum, they provide a practical framework for comparing experimental data with simulations that incorporate specific spectral changes. By analyzing these variables across both data and simulated scenarios, one can identify the spectral modifications that best reproduce the observed detector responses. In this way, \( R \) and \( \theta \) serve as effective intermediaries between detector-level observables and the underlying energy distributions of the shower components.

It should be noted that light collection in the WCD depends on several optical parameters, including the photomultiplier light collection efficiency, the attenuation of Cherenkov light in water, and the reflectivity of the Tyvek lining. These parameters are typically calibrated using the omnidirectional flux of atmospheric muons by correlating the observed signal peak with that of vertically traversing muons, commonly referred to as Vertical Equivalent Muons (VEM)~\cite{PierreAuger:2005znw}. While this procedure effectively calibrates the mean signal response, long-term monitoring has revealed that WCD optical parameters degrade over time, as indicated by changes in the ratio of the signal peak to its integrated charge~\cite{PierreAuger:2020flu}. One likely cause of this effect is the gradual deterioration of Tyvek reflectivity.

To evaluate the robustness of the proposed method against such ageing effects, we artificially reduced the Tyvek reflectivity by approximately 7\% to emulate the observed degradation, and then assessed its impact on the \( R \) and \( \theta \) observables after applying the standard VEM calibration procedure. As shown in Fig.~\ref{fig:graph_polar} (right), the changes observed relative to the nominal reflectivity were negligible. Furthermore, it is important to emphasize that any genuine modifications in the particle energy spectrum should produce consistent effects across all three detectors in the system, further reinforcing the interpretation that observed deviations arise from changes in shower characteristics rather than from detector calibration artifacts.

Further details and validation tests of the method presented in this section—particularly those focusing on the muonic shower component and its dependence on shower reconstruction parameters—can be found in~\cite{Assis:2025mis}.

\section{Energy spectrum from a hybrid station and the MPD}
\label{sec:MPD}

The Muon Production Depth (MPD) observable exhibits a rich phenomenology that has been extensively studied (see, for instance,~\cite{2012_Andringa_MPD}) and has been used to perform stringent tests of hadronic interaction models~\cite{PierreAuger:2014zay}. Its reconstruction relies on combining the shower geometry and the arrival time of the shower front with the arrival time of the detected muon~\cite{2012_Cazon_MuonTransport}. This approach allows us to pinpoint the muon's production point along the shower axis, provided the geometrical delay can be corrected for the kinematical delay. The latter depends on the energy spectrum of secondary muons and can significantly influence the inferred depth of the MPD profile~\cite{Espadanal:2016jse}, \( X^\mu_{\mathrm{max}} \), which is known to be sensitive to mass composition.

The height at which a muon is produced can be estimated using the following expression:

\begin{equation}
    z \simeq \frac{1}{2} \left( \frac{r^2}{c\left( t - \left< t_\varepsilon \right> \right)} - c \left( t - \left< t_\varepsilon \right> \right) \right) + \Delta - \left< z_\pi \right>
    \label{eq:mpd}
\end{equation}

where \( t \) is the muon's arrival time relative to the shower core, \( r \) is the distance from the shower axis, $\Delta$ accounts for the vertical displacement between the point on the ground where the muon is detected and the reference plane perpendicular to the shower axis passing through the shower core, and \( \left< z_\pi \right> \) is a small correction related to the pion decay altitude. The term \( \left< t_\varepsilon \right> \) represents the kinematical delay, typically obtained from Monte Carlo shower simulations.

As previously discussed, the kinematical delay term is sensitive to the muon energy. Therefore, if the arrival direction of the muon is known, equation~\ref{eq:mpd} could, in principle, be inverted to extract \( \left< t_\varepsilon \right> \), thus providing a handle on the muon energy distribution.

\begin{figure}[ht!]
    \centering
    \includegraphics[width=0.49\linewidth]{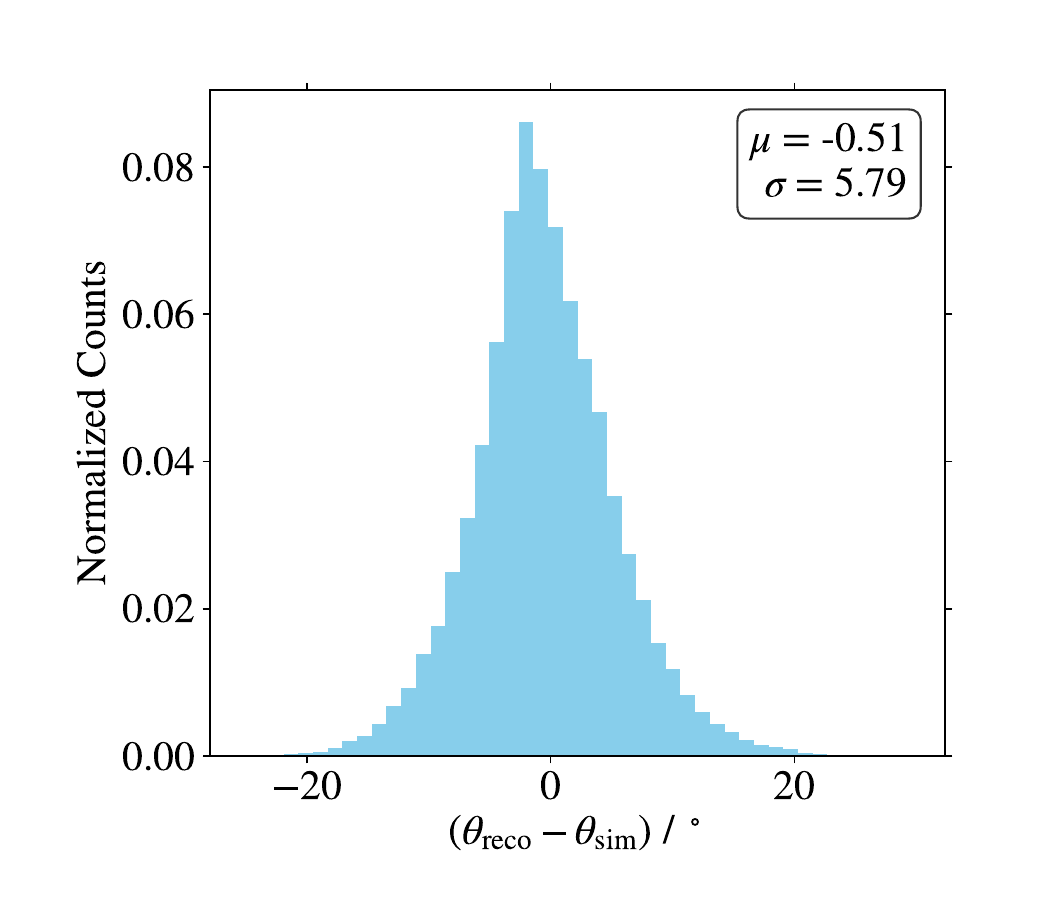}
    \includegraphics[width=0.44\linewidth]{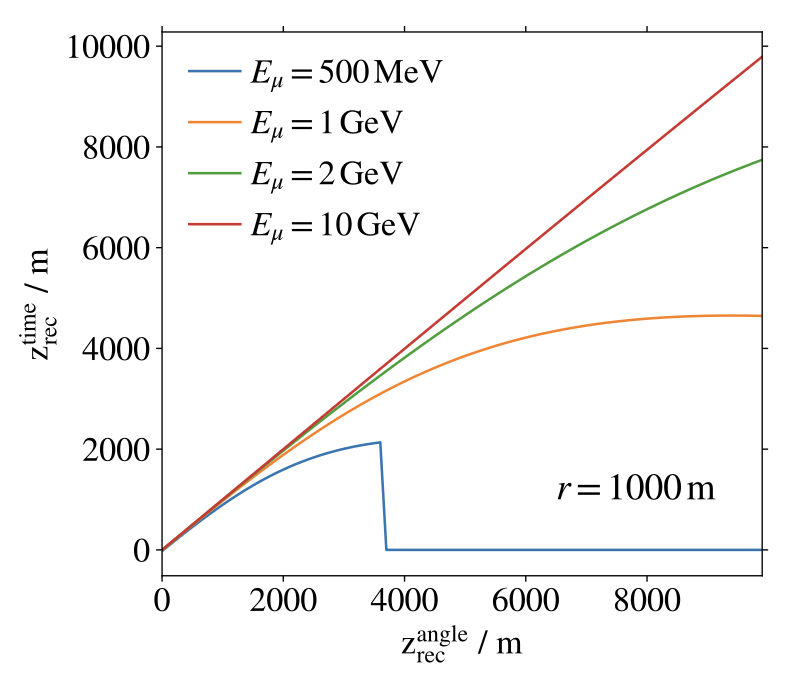}
    \caption{(left) Distribution of the residuals between the reconstructed and simulated muon zenith angles. (right) Correlation between the reconstructed production height obtained via the MPD algorithm, \( z^{\rm time}_{\rm rec} \), and that inferred from the muon arrival direction, \( z^{\rm angle}_{\rm rec} \), shown for different muon energies. Results correspond to a detector station located 1000\,m from the shower core.}
    \label{fig:mpd}
\end{figure}

As shown in~\cite{Alvarez-Muniz:2023hiu,Alvarez-Muniz:2025rct}, it is possible to reconstruct the incoming direction of a high-energy neutrino—which interacts nearby, producing a high-energy lepton—using a single water-Cherenkov detector (WCD), provided it is equipped with a sufficient number of photomultiplier tubes (PMTs), typically three at the bottom and one on top. The direction of the primary particle can be inferred by analyzing the PMT signal time traces with the aid of machine learning algorithms.

The MARTA station, consisting of a single WCD and RPCs positioned underneath, offers a setup analogous to the one described above. In this configuration, a muon traverses the WCD, producing a signal that is recorded by the three PMTs located at the top of the station, and subsequently hits a single pad of the RPC array, with an area on the order of some $\mathrm{cm}^2)$. This geometry enables precise timing and spatial measurements, facilitating directional reconstruction similar to the multi-PMT configuration.

For muons striking a specific RPC pad, the corresponding PMT signals are collected and analyzed using a Transformer-based neural network. The input consists of time traces from three photomultiplier tubes (PMTs), each represented as a one-dimensional array with 15 time bins capturing the muon-induced signal. The input tensor is first permuted so that each time bin is treated as a token in a sequence, and then passed through a linear embedding layer followed by a positional encoding step.

This sequence is subsequently processed by a stack of Transformer encoder layers, each employing multi-head self-attention with configurable parameters, including model dimension, number of attention heads, and feedforward network size. The output from the Transformer is flattened and, if desired, concatenated with additional input variables. The resulting vector is then passed through a series of fully connected (dense) layers with configurable widths, activation functions, and dropout rates. The final layer consists of a single neuron that outputs the predicted angle.

The model is trained to predict a continuous output value—namely, the incoming muon direction—based on the time structure of the PMT signals and any auxiliary input features provided.

Preliminary results from this approach—achieved by training the neural network with injected \( 1\,\mathrm{GeV} \) muons—are shown in Fig.~\ref{fig:mpd} (left). The figure demonstrates that an angular resolution of approximately \( 5^\circ \) can already be achieved with negligible bias. Although this result was obtained under idealized conditions, i.e., without electromagnetic contamination, it provides strong evidence for the feasibility of the proposed method.

Figure~\ref{fig:mpd} (right) illustrates the sensitivity of this approach using a simplified toy Monte Carlo for vertical showers, with a detector station located at a distance \( r = 1000\,\mathrm{m} \) from the shower core. In this test, the quantity \( z^{\mathrm{time}}_{\mathrm{rec}} \) is reconstructed using the MPD algorithm, while \( z^{\mathrm{angle}}_{\mathrm{rec}} \) is computed assuming the muon direction can be inferred using the neural network described above. The plot shows that this method offers reasonable sensitivity for muons with energies of \( \mathcal{O}(\mathrm{GeV}) \), particularly for those produced at high altitudes. Notably, an abrupt cutoff appears for lower-energy muons (around \( 500\,\mathrm{MeV} \)), which is attributed to energy losses through ionization -- feature that could also be exploited to perform more stringent tests of the shower muon energy spectrum.

\section{Summary}
\label{sec:summary}

In this work, we present two complementary strategies to access the energy spectra of the electromagnetic and muonic components of extensive air showers using a single hybrid detector station, composed of a surface scintillator detector, a water Cherenkov detector, and resistive plate chambers. The first method explores the different energy-dependent responses of the three detectors to individual particles, enabling sensitivity to modifications in the high-energy tail of the electromagnetic component at ground level. Signal correlations between detectors are analyzed using the estimators $R$ and $\theta$, which are shown to be robust against detector aging effects. The second method reconstructs the arrival direction of muons using PMT time traces in the WCD and RPC pad information, processed via a Transformer-based neural network. When combined with the Muon Production Depth framework, this allows for the extraction of the kinematical delay and, consequently, access to the energy spectrum of low-energy muons. These approaches open the door to a deeper understanding of hadronic interactions at ultra-high energies, while offering experimentally feasible solutions using compact, multi-purpose detection systems.

\section*{Acknowledgments}

This work has been funded by Fundação para a Ciência e Tecnologia, Portugal, under project \url{https://doi.org/10.54499/2024.06879.CERN}.

{\small
\bibliographystyle{JHEP}
\bibliography{references}}

\end{document}